\documentclass[11pt,english,twocolumn]{article}
\usepackage{mathptmx}
\usepackage{epstopdf} 
\usepackage{grffile}

\usepackage[T1]{fontenc}
\usepackage[latin9]{inputenc}
\usepackage{geometry}
\usepackage{amsthm}
\usepackage{amsmath}
\usepackage{graphicx}

\makeatletter
\numberwithin{equation}{section}
\numberwithin{figure}{section}

\usepackage{color}
\IfFileExists{lmodern.sty}{\usepackage{lmodern}}{}

\makeatother

\usepackage{babel}
\begin{document}

\title{Wave Propagation in 1-D Spiral geometry}

\author{Deep Chatterjee and K. Rajesh Nayak\\
CESSI and Dept. of Physical Sciences, IISER-Kolkata,\\ Mohanpur, West Bengal, India.}
\maketitle
\begin{abstract}
In this article, we investigate the wave equation in spiral geometry
and study the modes of vibrations of a one-dimensional\ (1-D) string
in spiral shape. Here we show that the problem of wave propagation
along a spiral can be reduced to Bessel differential equation and
hence, very closely related to the problem of radial waves of two-dimensional\ (2-D)
vibrating membrane in circular geometry. 
\end{abstract}



Our goal is to study the modes of a  string  bent
along a spiral. 
At the first glance, the problem appears to be only  of academic interest. 
However, it should be noted that the spiral geometries are  common in nature, 
specially those  related to wave propagations. The most
popular of spiral geometry, namely logarithmic spiral manifests itself
frequently in the nature in various forms. For example, simple conch to
arms of the majorities of galaxies are in spiral shape. Also, one of the 
most efficient waveguides in nature, the inner ear of mammals, is also  
a spiral! It has been shown that, the spiral shape of Basilar membrane
plays an important role in hearing \cite{Ear2006}. Here, we study the 
normal modes of vibration in this geometry and hope to understand its 
significance. 

Though there is vast amount of literature 
on the wave equation in standard geometries \cite{Allan,Arfken,Feshbach}, 
there has been little work in the spiral coordinate system, especially 
at the undergraduate level.
In our approach, we first do a coordinate
transform to reduce the wave equation to a Bessel differential equation.
This makes the problem similar to that of radial part of wave equation
in circular geometry or radial part of circular vibrating membrane.
The solutions are expressed in terms of the Bessel and Neumann
functions and in general, the modes are not integral multiples of  
fundamental mode. We investigate the solutions satisfying some of the 
standard boundary conditions.

\section{Spiral geometry and the wave equation}

\begin{figure}[h]
\begin{center}
\includegraphics[scale=0.495]{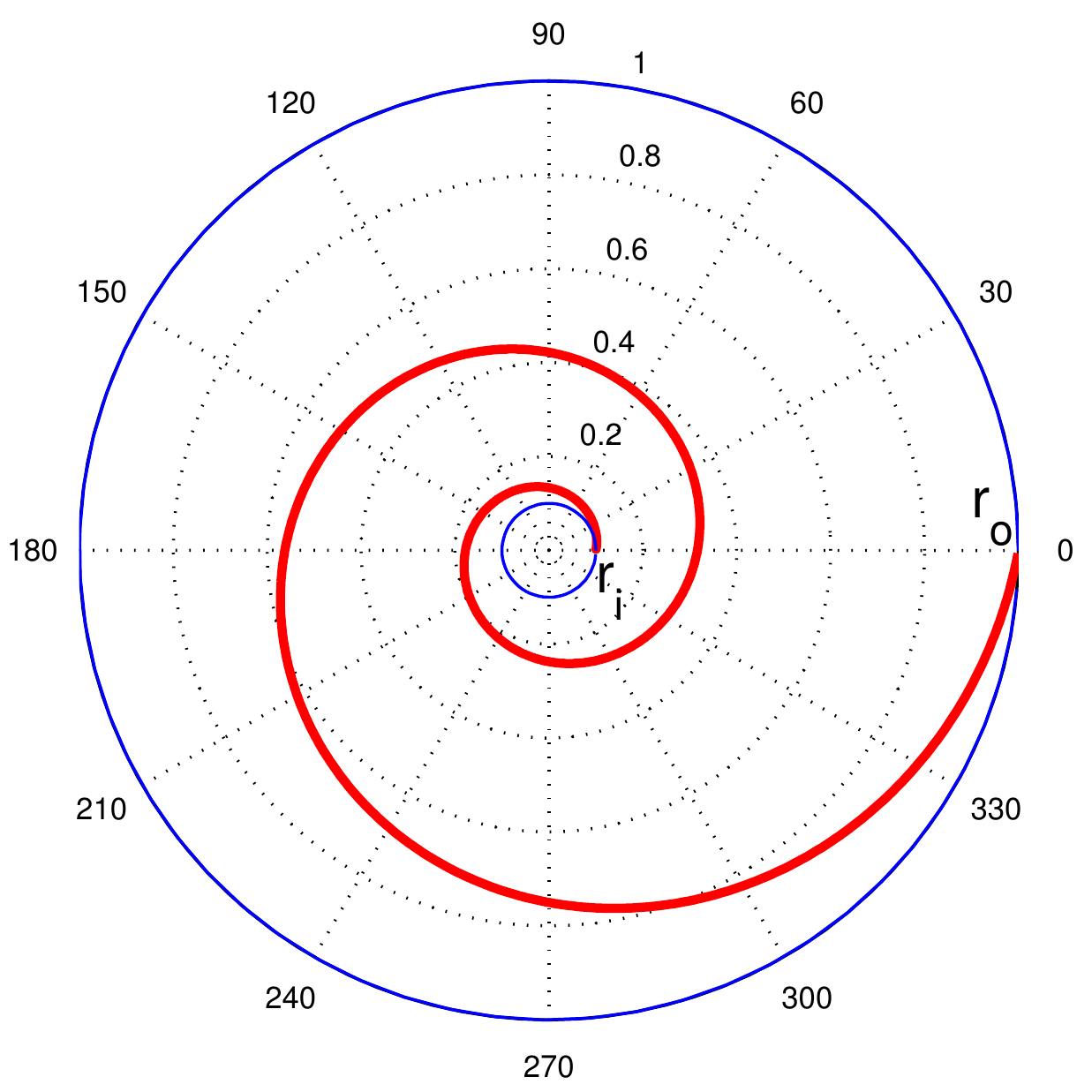}
\caption{\label{fig:spiral}Spiral starts at inner radius $r_i=0.1$ and reaches location $r_o=1$ after taking two turns.}
\end{center}
\end{figure}

In this section,  we discuss the geometry of logarithmic spiral. We would like
to have spiral connecting between $r_{i}$ to $r_{o}$ after taking
$n$ turns as shown in the Fig.~(\ref{fig:spiral}). The equation for logarithmic 
spiral is  given by:
\begin{equation}
r=r_{i}e^{\mu\theta}\,, \label{eq:sp1}
\end{equation}
where, $\mu$  describes how fast the spiral is accelerating  outwards. We would like to 
parameterize the spiral only in terms of $r_{i}$,
$r_{o}$ and $n$. Expressing $\mu$ in terms of these parameters, we have:
\begin{equation}
\mu=\frac{1}{2\pi n}\log\left(\frac{r_{o}}{r_{i}}\right)\,.\label{eq:wind}
\end{equation}
We see that, Larger number of turns, spiral slowly progresses outwards.
The length of the spiral is given by $s=\intop_{0}^{2n\pi}ds$, where $ds=\sqrt{r^{2}+\left(\frac{dr}{d\theta}\right)^{2}}\: d\theta$.
For logarithmic spiral we get:
\begin{equation}
s=\sqrt{\left(1+\frac{1}{\mu^{2}}\right)}\left(r_{o}-r_{i}\right)\,.
\end{equation}
Considering the relation between $\mu$ and $n$ as given in Eq.~(\ref{eq:wind}), we see that the length is proportional to  the 
number of turns $n$. 
Here, we could like to note that $n$ need not be integer, it  gives the measure of angle in terms of $2 \pi$
radians.

\begin{figure}
\begin{center}
\includegraphics[scale=0.45]{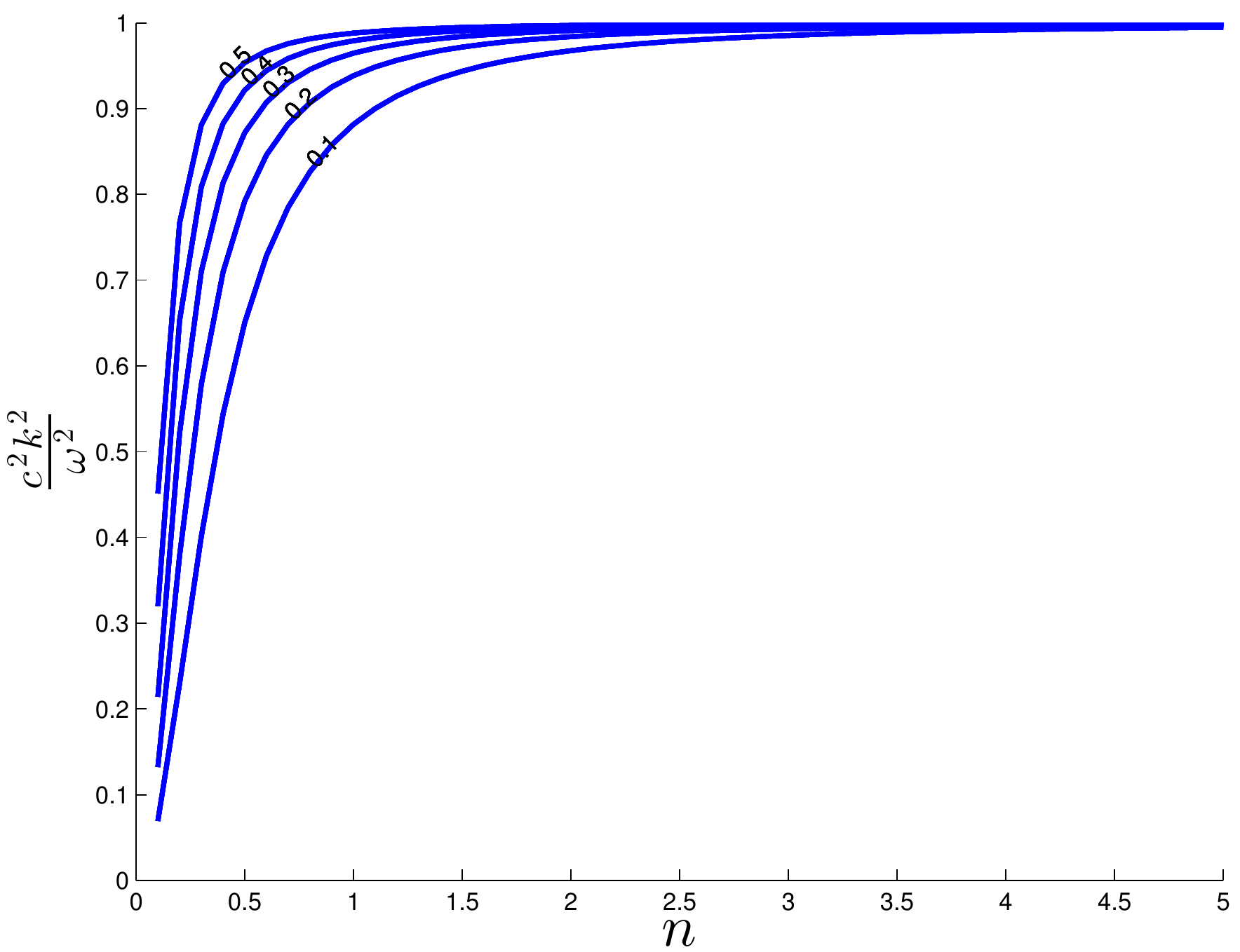}
\caption{Plot of dispersion relation as function of number of turns, '$n$'.}
\label{fig:disper}
\end{center}
\end{figure}

Next, we investigate the wave equation in the spiral geometry. 
We start with the scalar wave equation in the 2-D polar coordinates
$\left(r,\,\theta\right)$, given by:
\begin{equation}
\frac{1}{r}\frac{\partial}{\partial r}\left(r\frac{\partial\psi}{\partial r}\right)+\frac{1}{r^{2}}\frac{\partial^{2}\psi}{\partial\theta^{2}}-\frac{1}{c^{2}}\frac{\partial^{2}\psi}{\partial t^{2}}=0\,.
\end{equation}
We constrain this wave equation along a logarithmic spiral. Using Eq.~(\ref{eq:sp1}), we 
 directly  substituting for $\frac{\partial}{\partial\theta}$ and $\frac{\partial^{2}}{\partial\theta^{2}}$ as:
\begin{equation}
\frac{\partial^{2}\psi}{\partial\theta^{2}}=\mu^{2}r^{2}\frac{\partial^{2}\psi}{\partial r^{2}}+\mu^{2}r\frac{\partial\psi}{\partial r}\,.\label{eq:spiraleqapplied}
\end{equation}
After substitution and simplification, we obtain: 
\begin{equation}
r^{2}\frac{d^{2}}{dr^{2}}\psi\left(r\right)+r\frac{d}{dr}\psi\left(r\right)+k^{2}r^{2}\psi\left(r\right)=0\,,\label{eq:requation}
\end{equation}
where, 
\begin{equation}
k^{2}=\frac{\omega^{2}}{c^{2}\left(1+\mu^{2}\right)}\,.\label{eq:kvalue}
\end{equation}
 Here, for simplicity, we replace $\frac{\partial\psi}{\partial r}$
by $\frac{d\psi}{dr}$. From Eq.~(\ref{eq:kvalue}), we see that  the spatial frequencies are scaled a
factor of $\left(1+\mu^{2}\right)^{-\frac{1}{2}}$. 
The relation between $k$ and
$\omega$, the {\it dispersion relation}, clearly depends on the geometry
of the spiral. Plot of $\frac{c^2 k^2}{\omega ^2}$ as function of number 
of turns $n$ is shown in Fig.~(\ref{fig:disper}). When the
number of turn becomes large, the relation reduce to that in of a straight
string. However, when $n$ is small, we see the scaling of the spatial
frequencies. In the case of  linear string, 
the dispersion relation  leads to the conventional relation  $c=f \lambda$,
where  $\lambda$ is
wavelength, $f=\omega/2\pi$  is frequency and $c$  is the speed of the wave.
In spiral geometry, this relation needs to be corrected by 
a geometric factor as given by Eq.~(\ref{eq:kvalue}). Overall, for a 
 given frequency, wavelength is longer by a factor of $\sqrt{1+\mu^2}$,
compared to a linear-string.

The wave equation on the spiral, given by Eq.~(\ref{eq:requation}), is the
well known Bessel differential equation. The general 
solutions are given in terms of Bessel and Neumann functions as: 
\begin{equation}
\psi=A_{1}J_{0}\left(kr\right)+B_{1}N_{0}\left(kr\right)\,,
\end{equation}
 where $J_{0}$ and $N_{0}$,  are Bessel and Neumann function of order
zero respectively. However, we need to investigate the solutions with
physically important boundary conditions which may constrain solutions and frequencies.
In the next sub-section we study some of the boundary conditions.
One important point to notice here is the similarity in the solution along a spiral
and that in the radial wave equation in the  plane-polar geometry. In both we get the 
Bessel differential equation.

\section{Spiral string fixed at both ends \label{sub: Spiral-spring-with}}

In this section, we extend the classic linear vibrating string problem (see \cite{Allan} for extensive details)
with both end fixed to spiral geometry. Let the inner and outer ends
of a finite spiral string are fixed at $r=r_{i}$ and $r=r_{o}$, respectively.
Explicitly, the boundary conditions are given by $\psi\left(r=r_{i},\, t\right)=\psi\left(r=r_{o},\, t\right)=0$,
without loss of generality, the solution can be written as:
\begin{equation}
\psi=\sum_{n=1}A_{n}\left[\cos\delta\: J_{0}\left(\kappa_{n}r\right)+\sin\delta\: N_{0}\left(\kappa_{n}r\right)\right]\cos\omega_{n}t\,.\label{eq:expspi}
\end{equation}
Here, $\delta$ is chosen such that, the boundary condition at $r=r_{o}$
is satisfied, resulting in the condition:
\begin{equation}
\tan\delta=-\frac{J_{0}\left(\kappa_{n}r_{o}\right)}{N_{0}\left(\kappa_{n}r_{o}\right)}\,.\label{eq:ps}
\end{equation}
However, all values of $\kappa_{n}$ will not satisfy the boundary
condition at $r_{i}$ and the allowed values of $\kappa_{n}$ can
be determined from the equation:
\begin{equation}
\cos\delta\: J_{0}\left(\kappa_{n}r_{i}\right)+\sin\delta\: N_{0}\left(\kappa_{n}r_{i}\right)=0\,.\label{eq:bcroot}
\end{equation}
Numerical solutions for $\kappa_{n}$ for $n=1$ to $5$ is obtained for  few values of $r_i$, with fixed value of 
$r_o=1$ and are tabulated  in Table~\ref{table:nonlin}. Corresponding frequencies, $\omega_n$ can be obtained 
from the  relation given in Eq.~(\ref{eq:kvalue}). 
The eigenfrequencies play important role because they correspond to the resonance condition, giving the modes at which string vibrates. 
%
\begin{table}
\centering
\begin{tabular}{|l|l|l|l|l|l|l|}
\hline
$r_i$ & $r_o$ & $k_1$ & $k_2$ & $k_3$ & $k_4$ & $k_5$ \\
\hline
$0.1$ & $1.0$ &  3.313  &  6.857  & 10.37 &  13.88 &  17.38  \\
\hline
$0.2$ & $1.0$ &  3.815 &   7.785  &   11.73 &  15.67 &  19.60 \\
\hline
$0.3$ & $1.0$ &  4.412 &   8.932  &   13.43 &   17.92 &   22.42  \\
\hline
$0.4$ & $1.0$ &  5.183 &  10.44   &   15.68 &   20.92 &   26.16  \\
\hline
$0.5$ & $1.0$ &  6.246 &  12.54 &   18.83 &   25.12 &   31.40 \\
\hline
\end{tabular}
\caption{Roots of Eq.~(\ref{eq:bcroot}) for various values of $r_i$ and $r_o$.}
\label{table:nonlin}
\end{table}

\subsection{Finite Spiral string with vibrating ring\label{sub: Finite-Spiral-spring}}

Here, we consider the boundary condition such that, outer
end of a string is rigidly fixed at $r=r_{o}$, while the inner end
at $r=r_{i}$ is attached to a oscillating ring with frequency $\Omega$.   The corresponding
 boundary conditions can be written as $\psi\left(r=r_{i},\, t\right)=F\cos\Omega t$, with oscillating
circle 
and $\psi\left(r=r_{o},\, t\right)=0$. 
Any Solution,  can now be written terms of linear combination of Bessel and Neumann function as:
\begin{equation}
\psi=A\left[\cos\delta\: J_{0}\left(\kappa r\right)+\sin\delta\: N_{0}\left(\kappa r\right)\right]\cos\Omega t\,.\label{eq:splbc3}
\end{equation}
 The constants $A$ and $\delta$ need to be determined from the boundary
conditions. As in the previous case, from the boundary condition at $r_{o}$, i.e $\psi\left(r=r_{o},\, t\right)=0$,
we get: 
\begin{equation}
\tan\delta=-\frac{J_{0}\left(\kappa r_{o}\right)}{N_{0}\left(\kappa r_{o}\right)}\,.\label{eq:splph3}
\end{equation}
In addition, from the boundary condition at $r=r_{i}$, we get:
\begin{equation}
A=\frac{F}{\left[\cos\delta\: J_{0}\left(\kappa r_{i}\right)+\sin\delta\: N_{0}\left(\kappa r_{i}\right)\right]}\,.\label{eq:splam3}
\end{equation}
As expected, at resonance,   we will not be able to determine the  constant $A$, the  solution becomes undefined. 

\section{Conclusion\label{sec: Conclusions}}

Spiral geometry offers interesting solutions to the wave equation
with various physically important boundary conditions. Further interesting solutions maybe obtained by using different densities along the string, as is known for the case of Indian drums \cite{Tabala, Raman}. The connection between 2-D circular and 1-D spiral is important as wave equation
in circular geometry is well understood and one can apply some of the
implications immediately to the spiral geometry. Similarities between
circular drum head and spiral string is indeed a useful result since
the entire membrane can be replaced by a 1-D spring giving the same
vibration modes. This can come handy since the material required to
create a membrane can be done with by using the spiral. 

The more general case may be dealt by transforming from Cartesian coordinates to a coordinate system where constant curves are spirals \cite{Gil}. The solutions of 2-D wave equation in such a system can also be given in terms of Bessel
functions, which will be discussed else where. Since analytic solutions are
available, otherwise complete numerical analysis can be avoided for
the applications such as vibration isolator with spiral/helical suspension.
One the important application under consideration is the wave propagation
in the basilar membrane of mammals. It has already been shown that,
the spiral geometry of inner ears of mammals has important implications \cite{Ear2006},
this may be further extended using results from spiral coordinates.

\end{document}